\documentclass[preprint,amsmath,amssymb,aps, prb, twocolumn, 10pt, a4paper,superscriptaddress]{revtex4-2}

\usepackage[utf8]{inputenc}
\usepackage[english]{babel}
\usepackage[T1]{fontenc}

\usepackage{amsfonts}
\usepackage{amsthm}

\usepackage{bm}
\usepackage{csquotes}
\usepackage[normalem]{ulem}
\usepackage[table,dvipsnames]{xcolor}

\PassOptionsToPackage{hyphens}{url}
\usepackage[breaklinks,colorlinks=true]{hyperref}
\usepackage[capitalise]{cleveref}
\usepackage{graphicx}

\usepackage{physics}
\usepackage{mathtools}
\usepackage{xfrac}

\usepackage[inline]{enumitem}
\usepackage{placeins}
\usepackage[caption=false]{subfig}
\usepackage{simpler-wick}

\usepackage{algorithm}
\usepackage[noend]{algpseudocode}
\usepackage{tensor}
\usepackage{newfloat}

\usepackage{xspace}
\usepackage{gensymb}
\usepackage{booktabs}

\newcommand{\nele}{n}
\newcommand{\norb}{m}
\newcommand{\ndet}{N_D}

\newcommand{\nparam}{N_{P}}

\newcommand{\vacuum}{{0}}
\newcommand{\nstates}{K}
\newcommand{\ExcState}{\Psi}
\newcommand{\SDet}{\Phi}
\newcommand{\SDeTen}{\phi}
\newcommand{\OExcState}{\Gamma}
\newcommand{\SES}{S}
\newcommand{\HES}{H}
\newcommand{\EigenS}{\Xi}
\newcommand{\EigenE}{E}
\newcommand{\CostFunction}{\mathcal{L}}
\newcommand{\HQuad}{\mathcal H}
\newcommand{\SQuad}{\mathcal S}
\newcommand{\HessS}{{\tilde{\mathcal S}}}
\newcommand{\GradS}{{\mathcal T}}
\newcommand{\SRed}{{\mathbb S}}

\newcommand{\HessH}{{\tilde{\mathcal H}}}
\newcommand{\GradH}{{\mathcal G}}
\newcommand{\HRed}{{\mathbb H}}

\newcommand{\methodnameacronym}{\mbox{EIDOS}\xspace}

\newcommand{\methodnameexc}{Excited-state Exact Iterative Determinant-Orbital Solver\xspace}
\newcommand{\methodnameexcacronym}{\mbox{EXIDOS}\xspace}

\newcommand{\pma}[1]{{\begin{pmatrix} #1 \end{pmatrix}}}

\newcommand{\silentcite}[1]{{\begin{@fileswfalse}\cite{#1}\end{@fileswfalse}}}

\DeclareMathOperator{\Adj}{Adj}

\begin{document}

\title{Variational low-energy subspaces for chemically accurate excited states}

\author{Clemens Giuliani}
\email{clemens.giuliani@epfl.ch}
\affiliation{Institute of Physics, \'{E}cole Polytechnique F\'{e}d\'{e}rale de Lausanne (EPFL), 1015 Lausanne, Switzerland}
\author{Rocco Martinazzo}
\email{rocco.martinazzo@unimi.it}
\affiliation{Department of Chemistry, Universit\`a degli Studi di Milano, 20133 Milano, Italy}
\author{Giuseppe Carleo}
\email{giuseppe.carleo@epfl.ch}
\affiliation{Institute of Physics, \'{E}cole Polytechnique F\'{e}d\'{e}rale de Lausanne (EPFL), 1015 Lausanne, Switzerland}
\author{Riccardo Rossi}
\email{riccardo.rossi@epfl.ch}
\affiliation{Institute of Physics, \'{E}cole Polytechnique F\'{e}d\'{e}rale de Lausanne (EPFL), 1015 Lausanne, Switzerland}
\affiliation{CNRS, Laboratoire de Physique Th\'eorique de la Mati\`ere Condens\'ee, Sorbonne Universit\'e, 75005 Paris, France}

\begin{abstract}
Accurate electronic excited states are essential for photochemistry, spectroscopy and non-adiabatic molecular dynamics, but high-level calculations often scale steeply and require prior knowledge of the target state's character or symmetry. Here we show that variational excited-state optimization can be reformulated as an iterated ground-state-like problem for a low-energy subspace of the electronic Hamiltonian. Applying this variational principle to non-orthogonal Slater determinants leads to {\methodnameexcacronym}, an automatic method for excited state calculations controlled only by the number of states and determinants per state. {\methodnameexcacronym} optimizes multiple excited states simultaneously, without explicit orthogonality constraints or imposed spin and point-group symmetries. Benchmarks against FCI and state-of-the-art quantum chemistry methods show chemical accuracy for a multitude of states in N\textsubscript{2} and CO, charge-transfer states in HCl, Rydberg states in NH\textsubscript{3}, double excitations and extended potential-energy curves in C\textsubscript{2}, and avoided crossings and conical intersections in ethylene. These results establish {\methodnameexcacronym} as a low-scaling, fully variational route to chemically accurate excited states.

\end{abstract}

\maketitle

\section{Introduction}
Electronic excited states govern many of the processes at the heart of photochemistry, spectroscopy and non-adiabatic molecular dynamics. Predicting them accurately remains challenging for theoretical methods because the low-energy spectrum of a molecule can contain states of very different character, including valence, Rydberg, charge-transfer and doubly excited states, often separated by small energy gaps or connected through avoided crossings and conical intersections.

Existing excited-state methods offer different compromises between accuracy, scaling and generality. Time-dependent density-functional theory, coupled-cluster theory, algebraic diagrammatic construction, multireference perturbation theory, selected configuration interaction and quantum Monte Carlo techniques have all become important tools for molecular spectroscopy and photochemistry~\cite{Casida2012_tddft_review,Dreuw2005_headgordon_single_ref_exc_review,Dreuw2014_adc_review,Koch1990_cc_response_fun,Christiansen1995,Stanton1993_barlett_eom_cc,krylov2008eom,Purwanto2009_shiwei_c2_afqmc_exc_singlet,Sneskov2011_excited_state_cc_review,Dreuw2014_adc_review,lischka2018_multireference_excited_states_review,SerranoAndrs1993_casscf_caspt2_exc,angeli2001nevpt2,Holmes2017_shci_exc,Blunt2015_fciqmc,JimnezHoyos2013b_excited_states,Nite2019}. 
Yet calculations often remain state dependent: they require  specifying a symmetry sector, enforcing orthogonality to lower states, or choosing an ansatz suited to a presumed excitation character. Such choices are effective when the relevant states are known in advance, but can become fragile near degeneracies and conical intersections.

This points to the need for methods that optimize the low-energy spectrum as a whole rather than targeting states one at a time. Several non-orthogonal trial states are optimized together, and the Hamiltonian is diagonalized within their span. The resulting cost function is invariant under rotations of the trial states and, by the min--max theorem, yields variational upper bounds to the low-energy spectrum. This avoids explicit orthogonality constraints and treats all targeted states on equal footing. Although related subspace principles have recently been explored with neural-network wave functions~\cite{Pfau2024,ma_deng_2506.08594_trapped_ion_nqs,doug_2507.10287_wedge,adrien_2507.08930}, and tensor-network ans\"atze~\cite{leiwang_2504.21459}, existing approaches rely on fermionized copies of the original Hilbert space. While elegant and general, this construction substantially increases the cost relative to ground-state optimization and makes its implementation within quantum-chemical methods challenging.

Here we show that the low-energy subspace variational problem can be reformulated as an iterated ground-state optimization. At fixed values of all other trial states, the cost function becomes a generalized Rayleigh quotient for a single state, defined by an effective Hamiltonian and metric generated by the rest of the subspace. This observation turns excited-state optimization into a sequence of effective ground-state problems, creating a direct route to extend efficient ground-state wave-function optimizers to multiple excited states.

We combine this formulation with the recently introduced EIDOS approach for optimizing sums of non-orthogonal Slater determinants~\cite{paper1}. The resulting method, \methodnameexcacronym, represents each targeted state as a compact expansion of unconstrained, non-orthogonal determinants. It has only two systematic control parameters: the number of states and the number of determinants per state. It does not require assumptions about spin or spatial symmetry, nor about the character of the target excitations. Symmetries can instead be assigned a posteriori by evaluating the corresponding operators within the optimized low-energy subspace.
\begin{figure*}[t!]
\includegraphics{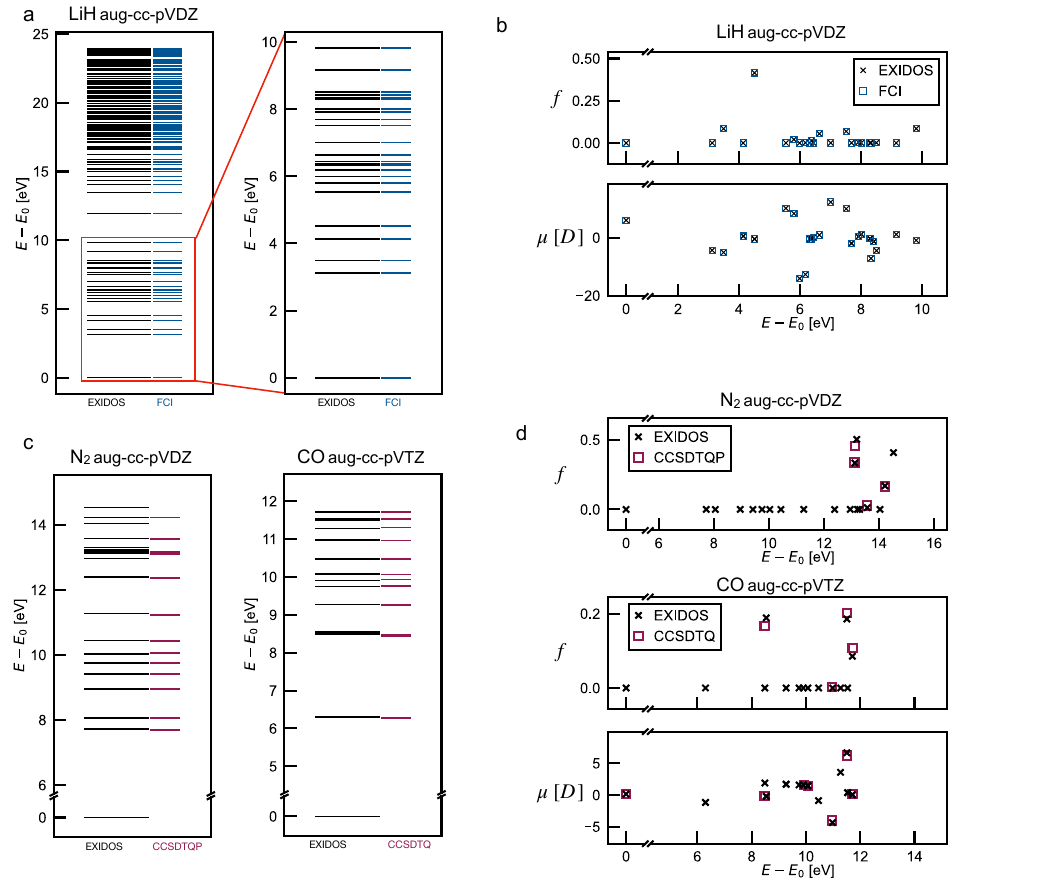}
\caption{{\it Automatic optimization of the low-energy subspace.} (a-b) {\methodnameexcacronym}(256,32) calculation of LiH spectrum, oscillator strength $f$ and dipole moment $\mu$, compared with FCI.
(c-d) All-electron {\methodnameexcacronym}(24, 256) and {\methodnameexcacronym}(33, 256) results for the N\textsubscript{2} and CO molecules, respectively, are compared with the benchmark results from Ref.~\cite{loos_jctc_2018_benchmark,Loos2025_questdb}. 
We note that panel (c-d) includes additional low-lying triplet states absent from the benchmark. 
}
\label{fig:method}
\end{figure*}

We validate EXIDOS across molecular benchmarks that probe dense spectra, different symmetries and distinct excitation mechanisms. The method reaches chemical accuracy against full configuration interaction for LiH and high-order coupled-cluster benchmarks for N\textsubscript{2} and CO, while using only the number of target states and determinants per state as inputs. The same framework recovers a posteriori the symmetries of the low-lying C$_2$ states along the dissociation curve, describes charge-transfer, doubly excited and Rydberg states in HCl, C\textsubscript{2} and NH\textsubscript{3}, and tracks the avoided crossing and conical intersections of distorted ethylene. These benchmarks establish EXIDOS as an automatic variational route to chemically accurate excited states.

\section{Results and discussion}

\begin{figure*}[t!]
\includegraphics{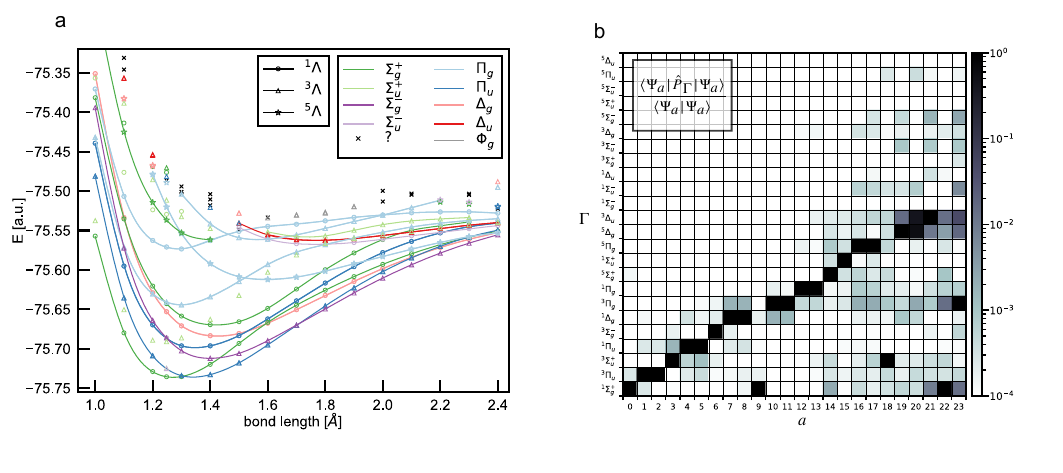}
\caption{
{\it (a) \methodnameexcacronym potential energy curve of the lowest states of C\textsubscript{2}}. We show an all-electron \methodnameexcacronym(24,256) calculation  in the aug-cc-pVDZ basis.
{\it (b) Symmetry retrieval from operator measurements.} We show the expectation value of the projector $\hat P_\Gamma$ onto each irreducible representation $\Gamma$ of a term symbol as function of the state index at equilibrium geometry for C\textsubscript{2} ($R = 1.248 \AA$), computed using the techniques described in Ref.~\cite{Scuseria2011,JimnezHoyos2012}.
}
\label{fig:c2}
\end{figure*}

\subsection{Excited states from an effective ground-state problem}
Traditional excited-state approaches often target individual eigenstates, using orthogonality constraints or predefined symmetry sectors. These choices can be effective, but require prior knowledge of the excitations and may introduce bias.

Here we instead optimize a low-energy subspace. Given $K$ linearly independent, not necessarily orthogonal, states $\{\ket{\Psi^a}\}_{a=0}^{K-1}$, we diagonalize $\hat H$ in their span. By the min-max theorem, the resulting Ritz energies are variational upper bounds to the lowest $K$ exact eigenvalues.
To improve these bounds, we optimize the span of these states by minimizing the following cost function~\cite{Pfau2024,ma_deng_2506.08594_trapped_ion_nqs,doug_2507.10287_wedge,adrien_2507.08930,leiwang_2504.21459}
\begin{equation}
\mathcal{L}
= \mathrm{Tr}(\hat P \hat H)
= \mathrm{Tr}[\SES^{-1}\HES],    
\end{equation}
where $\hat P$ projects onto the span of the states, and
$\SES_{ab}\coloneq\braket{\Psi^a}{\Psi^b}$, $\HES_{ab}\coloneq\mel{\Psi^a}{\hat H}{\Psi^b}$.
Because $\mathcal{L}$ depends only on the span, it treats all targeted states on equal footing and does not require explicit orthogonality constraints.

The key observation is that, when all states except $\ket{\Psi^a}$ are fixed, the same cost function becomes a generalized Rayleigh quotient,
\begin{equation}
\mathcal{L}
=
\frac{\mel{\Psi^a}{\hat H^{(a)}}{\Psi^a}}
{\mel{\Psi^a}{\hat S^{(a)}}{\Psi^a}},
\end{equation}
where the effective Hamiltonian $\hat H^{(a)}$ and metric $\hat S^{(a)}$ depend only on the remaining states. Thus, optimizing one state at fixed subspace environment is equivalent to a ground-state-like problem, {\it provided each state is variationally parametrized independently of the others}. Iterating this procedure over all states results in a self-consistent optimization of the full low-energy subspace. We leave the technical details of the derivation for the Methods section.

We realize this idea with sums of non-orthogonal Slater determinants. In {\methodnameexcacronym}$(K,N_D)$, each of the $K$ states is represented by $N_D$ independently parameterized determinants, in the spirit of the recently developed EIDOS method for ground-state calculations~\cite{paper1}.
\methodnameexcacronym retains the favorable $O(\norb^4)$ computational scaling of \methodnameacronym, offering a competitive cost relative to current state-of-the-art excited-state approaches.
The method constitutes a natural generalization of \methodnameacronym  to the simultaneous treatment of multiple states, reducing to the latter when K=1. After convergence, the physical excited states are obtained by diagonalizing the Hamiltonian in the larger space spanned by all optimized determinants.

\begin{figure*}[ht!]
\includegraphics{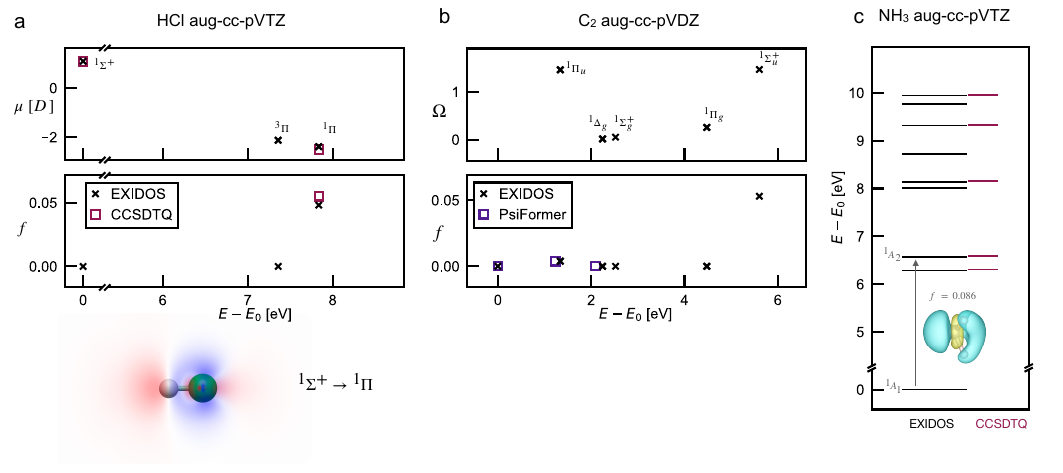}
\caption{\textit{Excited states with different electronic character.}
(a) Charge-transfer excitation in HCl. A frozen-core {\methodnameexcacronym}(5,256) calculation reproduces the excitation energy, dipole moments $\mu$ and oscillator strength $f$ of the 
${{}^1\Sigma^+ \rightarrow {}^1\Pi}$ transition, compared with CCSDTQ~\cite{Chrayteh2020_dip_benchmark}.
Our calculation also includes the lowest ${}^{3}\Pi$ state absent from the benchmark.
The density difference plot illustrates the charge redistribution for the ${{}^1\Sigma^+ \rightarrow {}^1\Pi}$ transition.
(b) Double excitations in C\textsubscript{2} at equilibrium. An all-electron {\methodnameexcacronym}(24,256) calculation recovers the low-lying singlet states with strong double-excitation character.
$\Omega$ quantifies the doubly excited character of each excited state: $\Omega \to 0$ indicates a state that cannot be described by single excitations~\cite{doCasal2023_trdm_double_excitation}. The Psiformer result is from Ref.~\cite{Pfau2024}, and for clarity of presentation we only show singlets.
(c) Rydberg states in NH\textsubscript{3}. A frozen-core {\methodnameexcacronym}(14,256) run reproduces the excitation spectrum when compared to CCSDTQ~\cite{loos_jctc_2018_benchmark,Loos2025_questdb}; The highlighted bright transition ($f = 0.086$) represents a diffuse Rydberg excitation, as shown by the density difference plot.
}
\label{fig:diverse-states}
\end{figure*}

\subsection{Benchmarking automatic low-energy subspace retrieval on LiH, N\textsubscript{2} and CO}
We first benchmark {\methodnameexcacronym} on LiH, N\textsubscript{2} and CO, using only the number of target states \nstates\ and the number of determinants per state $\ndet$ as inputs. These systems provide stringent comparisons with full configuration interaction and high-order coupled cluster, while probing dense spectra, near-degeneracies and optically active transitions.

LiH tests the numerical stability of the subspace optimization. In Fig.~\ref{fig:method}(a-b) {\methodnameexcacronym} reproduces the full-CI spectrum within less than $20$ meV using $32$ determinants per state, with a maximal deviation of less than $2$ meV for the first 35 states.
A single subspace calculation retrieves the lowest 256 states and gives oscillator strengths and dipole moments in agreement with FCI.

In Fig.~\ref{fig:method}(c-d) we compare {\methodnameexcacronym} spectra for N\textsubscript{2} and CO with state-of-the-art excited-state benchmarks. 
{\methodnameexcacronym} excitation energies are typically below chemical accuracy when compared to high-order coupled-cluster benchmarks, with an average absolute deviation of {$15$ meV} ($18$ meV) and a maximal absolute deviation of {$36$ meV} ($47$ meV) for N\textsubscript{2} (CO) for the states shown in the figure. {\methodnameexcacronym} consistently outperforms CC3 in these molecules, despite a lower asymptotic scaling in the number of basis-set orbitals. Oscillator strengths and dipole moments are obtained directly from the variational states, and they show good agreement with coupled cluster results at the quadruple and pentuple excitation level.

These benchmarks show that {\methodnameexcacronym} can recover chemically accurate low-energy spectra in a single calculation without prior assumptions about the target states. The accuracy is controlled systematically by \nstates\ and $\ndet$, and the same optimized subspace provides energies and transition properties. 
We also stress the crucial use of unconstrained Slater determinants, which provide an unbiased variational description of electronic correlation, without distinguishing a priori between static and dynamic contributions. Both are instead recovered adaptively and systematically as required by the system and the size of the determinant expansion.

\begin{figure*}[t!]
\includegraphics{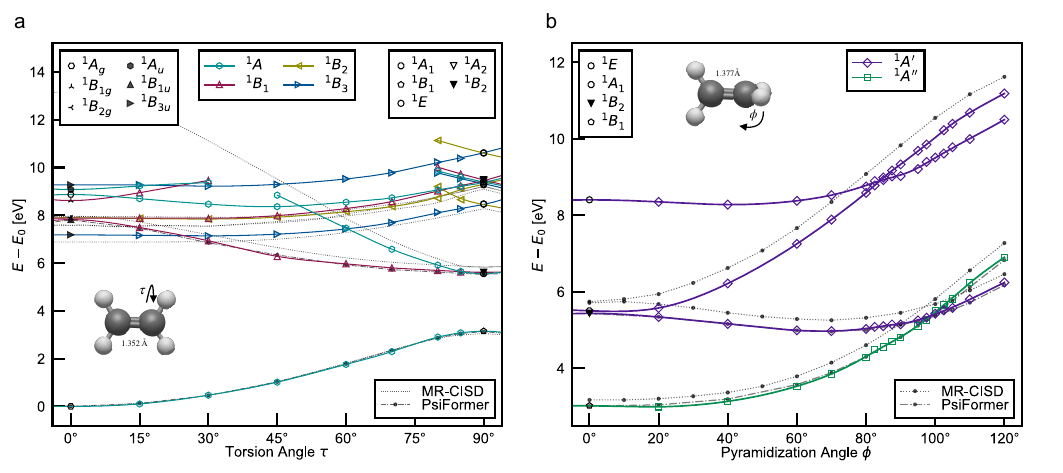}
\caption{
{\it (a) Bond torsion of the ethylene molecule starting from the planar geometry.}
{\it (b) Bond pyramidization of the ethylene molecule starting from the twisted-orthogonal geometry. }
 Results are obtained from an all-electron \methodnameexcacronym simulation of $\nstates=10$ states ($16$ for a few geometries) with $\ndet=256$ determinants in the aug-cc-pVDZ basis, and the geometries are from Ref.~\cite{Barbatti2004_mrci_ethylene}. 
We compare with MR-CISD in the same basis~\cite{Barbatti2004_mrci_ethylene}
and with PsiFormer~\cite{Pfau2024}. In both panels we use a $\hat S^2$ spin penalty to only select singlets in order to reduce the computational cost.
}
\label{fig:ethylene_trsn}
\label{fig:ethylene_pyrm}

\end{figure*}

\subsection{State-symmetry  restoration along the C\textsubscript{2} potential-energy curve}

Diatomic carbon is a challenging test case because its low-energy spectrum contains closely spaced singlet, triplet and quintet states of different spatial symmetries, together with strong multireference and double-excitation character. We use C\textsubscript{2} to test whether \methodnameexcacronym can recover this manifold without targeting individual symmetry sectors.

\Cref{fig:c2}(a) shows an all-electron \methodnameexcacronym calculation of the lowest 24 states along the C\textsubscript{2} potential-energy curve in the aug-cc-pVDZ basis. Unlike approaches that compute one or two states within selected symmetry sectors~\cite{Purwanto2009_shiwei_c2_afqmc_exc_singlet,Blunt2015_fciqmc,Sharma2015_dmrg_c2_symm,Holmes2017_shci_exc}, \methodnameexcacronym optimizes the full low-energy subspace at each geometry in a single calculation.

The symmetry recovery is quantified in \cref{fig:c2}(b) at the equilibrium geometry. For each optimized state, we evaluate the projector onto the symmetry sector associated with each term symbol. The resulting matrix has a dominant entry for each column, which is used to assign a term symbol to the state. The optimized states thus recover well-defined spin and point-group symmetries a posteriori. This is a nontrivial outcome, since no spin or spatial symmetry constraints are imposed during the variational optimization, beyond fixing the numbers of spin-up and spin-down electrons (see \cref{sec:uhf_ansatz}). The residual weight in other symmetry sectors is mainly confined to the highest states in the targeted manifold, where nearby states outside the optimized subspace can mix more strongly with states not included in the manifold. This residual mixing can be reduced systematically by increasing the number of targeted states $K$ to include those states.

\subsection{Capturing charge transfer, doubly excited, and Rydberg states}
We investigate whether \methodnameexcacronym can describe excited states with qualitatively different electronic character without modifying the ansatz or targeting specific transitions. \Cref{fig:diverse-states} illustrates three representative cases: a charge-transfer excitation in HCl, double excitations in C\textsubscript{2} and Rydberg states in NH\textsubscript{3}.

In HCl, the lowest ${}^{1}\Pi$ excitation has pronounced charge-transfer character, reflected in a reversal of the molecular dipole moment along the bond axis. As shown in \cref{fig:diverse-states}(a), \methodnameexcacronym reproduces both the excitation energy and the large change in dipole moment in agreement with high-order coupled-cluster benchmarks. The oscillator strength for the spin-allowed ${}^1\Sigma^+ \rightarrow {}^1\Pi$ transition is also in good agreement with CCSDTQ.

Diatomic carbon probes a different limitation of many excited-state methods: low-lying states with strong double-excitation character. In particular, the ${}^{1}\Sigma_g^+ \rightarrow {}^{1}\Delta_g$ and ${}^{1}\Sigma_g^+ \rightarrow {}^{1}\Sigma_g^+$ transitions have vanishing single-excitation fraction in CC3~\cite{Loos2019_double_excitations}. \Cref{fig:diverse-states}(b) shows that these states are correctly identified as doubly-excited by \methodnameexcacronym using the $\Omega$ metric of Ref.~\cite{doCasal2023_trdm_double_excitation}. 

Finally, we consider Rydberg excitations in NH\textsubscript{3}, which require a diffuse electronic description and are sensitive to the quality of the excited-state wave function. In the aug-cc-pVTZ basis, \methodnameexcacronym reproduces the low-energy Rydberg spectrum in close agreement with CCSDTQ, showing an average absolute deviation of $18$ meV and a maximal one of $25$ meV, see \cref{fig:diverse-states}(c). For the first bright transition we find an oscillator strength of $f=0.086$, and the corresponding difference density plot features the expected diffuse Rydberg character.
We remark that here we use the basis from previous benchmarks \cite{loos_jctc_2018_benchmark}, while a more accurate description of these Rydberg states would require more optimized basis sets \cite{2510.26751_levi_nn_shci}.

\subsection{Conical intersections and avoided crossings in ethylene}
Ethylene is a standard benchmark for excited-state methods because its low-lying spectrum combines several states of similar energy, mixed valence--Rydberg character and strong geometry dependence~\cite{Barbatti2004_mrci_ethylene,Feller2014_ethylene_cc_ci_review,Schmerwitz2022_levi_ethylene_dft,Entwistle2023,Pfau2024,Saade2024_ss_casscf_ethylene}. Starting from the planar molecule, twisting the C=C bond produces avoided crossings between the two lowest singlet states of the same symmetry, while subsequent pyramidalization of one CH$_2$ group leads to conical intersections. 
To focus on the singlet states involved in the conical intersections and the avoided crossings, we add an $\hat S^2$ penalty term that shifts triplet states out of the targeted low-energy window and decreases the computational cost. 

We  follow the torsion coordinate used in Ref.~\cite{Barbatti2004_mrci_ethylene}, rotating one CH$_2$ group around the C=C bond from the planar geometry to $\tau=90^\circ$, and change the point group from $D_{2h}$ to $D_2$ and finally to $D_{2d}$ at 90\degree. 
Across this path, \methodnameexcacronym follows the low-energy states smoothly and reproduces avoided crossings in the singlet manifold in  agreement with the MR-CISD and PsiFormer results (see \Cref{fig:ethylene_pyrm}(a)). We do not observe the spurious discontinuities that can affect state-specific optimizations in this geometry~\cite{Saade2024_ss_casscf_ethylene}.

Starting from the twisted-orthogonal structure, we next pyramidalize one CH$_2$ group up to $\phi=120^\circ$, during which the point group changes from $D_{2d}$ to $C_s$. As shown in \cref{fig:ethylene_pyrm}(b), \methodnameexcacronym reproduces the symmetry allowed crossing between the lowest ${}^1A'$ and ${}^1A''$ states at $\phi\simeq98.6^\circ$, consistent with previous MR-CISD and variational Monte Carlo results~\cite{Barbatti2004_mrci_ethylene,Entwistle2023,Pfau2024}. The calculation also resolves the conical intersection within the ${}^1A'$ manifold around $\phi\simeq84^\circ$, which is particularly challenging to obtain with symmetry-targeting techniques. 

\section{Conclusions}
We have introduced \methodnameexcacronym, a variational approach for computing molecular excited states by optimizing a low-energy subspace rather than individual eigenstates. The central result is that the subspace variational principle can be recast as a sequence of effective ground-state problems, allowing ground-state optimization strategies based on non-orthogonal Slater determinants to be extended directly to excited states. The method is controlled by only two systematic parameters, the number of targeted states \nstates\ and the number of determinants per state $\ndet$, and does not require orthogonality constraints, imposed spin or spatial symmetries, or prior knowledge of the excitation character.

Across the benchmarks considered here, this formulation gives chemically accurate low-energy spectra and transition properties for states of markedly different character. \methodnameexcacronym recovers dense spectra in LiH, N\textsubscript{2} and CO, assigns symmetries a posteriori along the C\textsubscript{2} potential-energy curve, and captures charge-transfer, doubly excited and Rydberg states without changing the ansatz. It also remains stable in geometry-dependent problems, reproducing the avoided crossing and conical intersections of ethylene while following the relevant states continuously through changes in molecular symmetry. These results show that optimizing the low-energy subspace as a whole provides a robust route to excited states in regimes where state-specific approaches can become sensitive to the choice of target, symmetry sector or initial guess.

The present implementation uses compact expansions of unconstrained Slater determinants, which can adaptively recover static and dynamic correlation but are not size-consistent in the strict asymptotic sense. This limitation suggests several natural extensions. The optimized \methodnameexcacronym subspaces could provide high-quality trial spaces for projector or Monte Carlo approaches, multi-Slater-Jastrow wave functions, or perturbative corrections~\cite{pineda_2019_slater_jastrow_excited,Schautz2004_filippi_dmc_exc_slaterjastrow,Burton2020_nocipt2,Mahajan2026_sharma_shiwei_afqmc_exc_as_gs}. 

\section{Methods}
\subsection{Subspace variational principle for excited states}\label{sec-var-theorem-exc-states}
\label{sec:var_principle}
In this section we review the generalization of the ground-state variational principle to excited states involving the subspace excited-state cost function. The approximation to the eigenstates of the Hamiltonian one obtains from the Hamiltonian are individually variational: they bound the energy of the $a$-th excited state from above.

Let $\{\ket{\ExcState^{a}}\}_{a=0}^{\nstates-1}$ be a set of quantum states, in general assumed to be non-orthogonal. We diagonalize the Hamiltonian $\hat H$ in the subspace spanned by the states $\{\ket{\ExcState^{a}}_{a=0}^{\nstates-1}\}$: this is equivalent to solve the following generalized eigenvalue problem (GEV)
\begin{equation}
\label{eq:eigval_problem_var_principle}
    H c_a = \epsilon_a\, S c_a,\qquad a\in\{0,\dots,\nstates-1\}
\end{equation}
where $\HES, \SES\in \mathbb C^{\nstates\times \nstates}$ are the Hamiltonian and the overlap matrices
\begin{equation}
\label{eq:mel_S_H}
\begin{aligned}
\SES_{ab} &\coloneq \braket{\ExcState^{a} }{\ExcState^{b}},\\
\HES_{ab} &\coloneq \mel{\ExcState^{a}}{\hat H}{\ExcState^{b}},
\end{aligned}
\end{equation}
and where $c_a\in\mathbb{C}^\nstates$ is the solution of the GEV. We assume here that $\SES$ is non-singular, which is equivalent to saying that the span of the vectors $\{\ket{\psi_a}\}_{a=0}^{K-1}$ is $K$-dimensional.

The eigenvectors $c_a \in \mathbb C^\nstates$ can be used to build a set of  $\nstates$ orthogonal states, $\{\ket*{\OExcState^{a}}\}_{a=0}^{\nstates-1}$
\begin{equation}
\label{eq:gamma_states}
    \ket*{\OExcState^{a}} \coloneq \sum_{b=0}^{\nstates-1} c_{ab}\, \ket{\ExcState^{b}}, \qquad \braket{\OExcState^{a}}{\OExcState^{b}}\propto \delta_{a,b}
\end{equation}
with average energy
\begin{equation}
\label{eq:gamma_energy}
    \epsilon_a = \frac{\mel*{\OExcState^{a}}{\hat H}{\OExcState^{a}}}{\braket*{\OExcState^{a} }{\OExcState^{a} }}.
\end{equation}

Let $\EigenE_{0} < \EigenE_{1} < \dots$ be the eigenvalues of $\hat H$, which we suppose non-degenerate for simplicity of presentation, and let $\ket{\EigenS^a}$ be the eigenstate of $\hat H$ associated to $\EigenE_a$.
In Ref.~\cite{leiwang_2504.21459,doug_2507.10287_wedge,adrien_2507.08930} it was pointed out that the min-max theorem provides the following important result
\begin{equation}\label{eq-minmax-theorem-bound}
    \EigenE_a \leq \epsilon_a,
\end{equation}
which means the average energy of the $a$-th state $\ket*{\OExcState^a}$ variationally bounds the energy of the $a$-th eigenstate $\ket{\EigenS^a}$ of the Hamiltonian.

Following Ref.~\cite{Pfau2024}, we introduce the loss function $\CostFunction$
\begin{equation}\label{eq-cost-function}
\CostFunction \coloneq \Tr\left [\SES^{-1}\, \HES \right].
\end{equation}
Using Eq.~\eqref{eq-minmax-theorem-bound}, one has
\begin{equation}
    \sum_{a=0}^{\nstates-1} E_a\le \CostFunction.
\end{equation}
The equality in the last equation is achieved if and only if
\begin{equation}\label{eq-span-equality}
    \textrm{Span}\big(\ket*{\EigenS^0},\dots,\ket*{\EigenS^{\nstates-1}}\big)=    \textrm{Span}\big(\ket*{\ExcState^{0}},\dots,\ket*{\ExcState^{\nstates-1}}\big).
\end{equation}

This provides what we call the subspace variational principle for excited states: if the wave function span of the vectors $\{\ket{\ExcState^a}\}_{a=0}^{\nstates-1}$  is chosen in some manifold and optimized to minimize $\mathcal{L}$, one can obtain an approximation to the first $\nstates$ excited states that is individually variational for each state (see Eq.~\ref{eq-minmax-theorem-bound}).

\subsection{Reducing the excited-state problem to an iterated ground-state optimization}\label{sec-effective-hamiltonian-state}

In the following we show that the cost function of \cref{eq-cost-function}, when viewed as a function of one of the states $\ket{\ExcState^{a}}$, can be written as a generalized Rayleigh quotient involving an effective Hamiltonian and an effective overlap operator. The variational excited-state optimization is then equivalent to an effective ground-state problem when viewed as a function of one state.

We rewrite the cost function of \cref{eq-cost-function} as
\begin{equation}\label{eq-loss-pfau-form-det}
\mathcal L = \Tr[\SES^{-1}\HES] = \frac{\Tr[\Adj(\SES) \HES]}{\det[\SES]}
\end{equation}
where $\Adj$ denotes the adjugate.
The determinant in the numerator can be written as
\begin{equation}
\label{eq:dets_expansion}
 \det[\SES] = \sum_{\sigma \in S_K} \mathrm{sgn}(\sigma) \prod_{a=0}^{\nstates-1} \braket*{\Psi^{a}}{\Psi^{\sigma[a]}},
\end{equation}
with $S_K$ being the symmetric group.
Similarly, for the numerator of \cref{eq-loss-pfau-form-det} we have that
\begin{equation}
\begin{aligned}
\label{eq:tradj_expansion}
    &\Tr[\Adj(\SES) \HES] =
    \\
    &\sum_{\sigma \in S_K} \mathrm{sgn}(\sigma) \sum_{b=0}^{\nstates-1}    \mel*{\Psi^{b}}{\hat H}{\Psi^{\sigma[b]}} \prod_{a=0, a\neq b}^{\nstates-1} \braket*{\Psi^{a}}{\Psi^{\sigma[a]}}.
\end{aligned}
\end{equation}
From \cref{eq:dets_expansion,eq:tradj_expansion} we can clearly see that, for any state $\ket{\ExcState^{a}}$, both numerator and denumerator are quadratic
\begin{equation}
\label{eq:ratio_quad_form_op}
    \mathcal L = \Tr[\SES^{-1}\HES] = \frac{
    \mel{\ExcState^{a}}{\hat H^{(a)}}{\ExcState^{a}}
    }{
    \mel{\ExcState^{a}}{\hat S^{(a)}}{\ExcState^{a}}
    },
\end{equation}
where $\hat H^{(a)}$ and $\hat S^{(a)}$ are some effective operators which do not depend on $\ket{\ExcState^{a}}$, but only on $\ket{\ExcState^{b}}$ for $b\neq a$.
We derive their full expressions in  \cref{appendix:derivation_operators}. We stress that it would be incorrect to state that the effective ground-state problem is equivalent to optimize one excited state at a time, as the true variational excited states are linear combinations of the $\ket{\ExcState^a}$ states, see \cref{eq:gamma_states}.

\subsection{Exact optimization of a linear ansatz for one state}
\label{sec-linear-optimization}
In this section we use the fact that $\CostFunction$, when viewed as a function of a single state, is equivalent to an effective ground-state problem to optimize exactly a single state. Mathematically, we use the fact that the cost function is the ratio of two quadratic forms (See~\cref{sec-effective-hamiltonian-state}) to show that a single linearly-parametrized excited state can be exactly optimized.

We now consider the following set of states
\begin{equation}\label{eq-one-state-ansatz}
    \Big\{\ket{\ExcState^{0}(\bm \theta)},\ket{\ExcState^1} ,\dots, \ket{\ExcState^{\nstates-1}}\Big\},
\end{equation}
where $\bm \theta\in\mathbb{C}^{N_P}$, $N_P$ is the number of variational parameters, and $\ket{\Psi^1},\dots, \ket{\Psi^{K-1}}$ are $\bm\theta$-independent fixed states.

Using Eq.~\eqref{eq-span-equality}, we point out that $\CostFunction$ is a function of the span of the vectors, which means that we can perform linear non-singular transformations of the vectors without changing the loss function. In particular, we can permutate the order of the states. Therefore, this justifies why we focus, without loss of generality, on optimizing the first state.

The scope of this section is the exact optimization of a state $\ket{\ExcState^{0}(\bm \theta)}$ that is linearly parametrized
\begin{equation}\label{eq-linear-ansatz}
\ket{\ExcState^{0}(\bm \theta)} = \sum_{p=1}^{\nparam} \theta_{p} \ket{\ExcState^{0}_{p}},
\end{equation}
where $\ket{\ExcState^{0}_{p}}$ do not depend on the parameters $\bm \theta$.

Following the variational theorem for excited states introduced in Sec.~\ref{sec-var-theorem-exc-states},  we wish to minimize the cost function $\CostFunction(\bm \theta)\coloneq \Tr\left [\SES(\bm\theta)^{-1}\, \HES(\bm\theta) \right]$, as defined in Eq.~\eqref{eq-cost-function}, with respect to the parameters $\bm \theta$. For the special case of the linear ansatz considered here, Eq.~\eqref{eq-linear-ansatz}, this can be done exactly, as we now illustrate. As proven in \cref{eq:ratio_quad_form_op} in the previous section, the cost function is a ratio of two quadratic functions of the state $\ket{\ExcState^{0}(\bm \theta)}$, which implies that we can write (using the linearity of Eq.~\eqref{eq-linear-ansatz})
\begin{equation}
\label{eq:tr_quad_form}
 \CostFunction(\bm \theta)=   \Tr[\SES^{-1}(\bm \theta)\,\HES(\bm \theta)] = \frac{{\bm\theta}^\dagger \HQuad \bm \theta}{{\bm\theta}^\dagger \SQuad \bm\theta},
\end{equation}
where $\HQuad,\SQuad \in \mathbb C^{\nparam \times \nparam}$ are the matrix elements of $\hat H^{(0)}$ and $\hat S^{(0)}$ (see \cref{eq:ratio_quad_form_op}) with the wavefunctions $\ket{\ExcState^{0}_p}$,
${\SQuad_{pq} =  \mel{\ExcState^{0}_p}{\hat S^{(0)}}{\ExcState^{0}_q}}$,
${\HQuad_{pq} =  \mel{\ExcState^{0}_p}{\hat H^{(0)}}{\ExcState^{0}_q}}$, given by

\begin{align}
\label{eq:squad_def}
\SQuad &= \HessS - \GradS^\dagger\, \SRed^{-1}\, \GradS,    \\
\label{eq:hquad_def}
\HQuad &= \HessH  - \GradS^\dagger \SRed^{-1} \GradH
-\GradH^\dagger \SRed^{-1} \GradS
+ \Tr[\SRed^{-1} \HRed]\, \SQuad\\
&\quad+ \GradS^\dagger \SRed^{-1} \HRed\SRed^{-1} \GradS \notag
\end{align}

where $\SRed$ ($\HRed$) is the $\nstates{-}1\times \nstates{-}1$ matrix obtained by removing the first row and column from $\SES$ ($\HES$) defined in \cref{eq:mel_S_H}, and $\GradS$,$\GradH$ are matrices of size $\nstates{-}1 \times \nparam$
\begin{align}
\label{eq:mel_grad}
    &\GradS_{ap} =  \braket{{\ExcState^{a}}}{\ExcState^{0}_p}, \quad
    &\GradH_{ap} =  \mel{{\ExcState^{a}}}{\hat H}{\ExcState^{0}_p}
\end{align}
    for $a\in\{1,\dots,{\nstates-1}\}$, and where
\begin{align}
\label{eq:mel_hess}
    &\HessS_{pq} =   \braket{\ExcState^{0}_p}{{\ExcState^{0}_q}}, \quad
    &\HessH_{pq} = \mel{\ExcState^{0}_p}{\hat H}{{\ExcState^{0}_q}}
\end{align}
are of size $\nparam \times \nparam$. The derivation of these formulae is presented in \cref{appendix:derivation_tensors}.

From \cref{eq:tr_quad_form}, one can immediately deduce that the cost function $\CostFunction(\bm \theta)$ is minimized  by a value of $\bm \theta$ satisfying the following generalized eigenvalue problem
\begin{equation}
  \label{eq:gev}
\HQuad \bm \theta = \varepsilon\, \SQuad \bm\theta,
\end{equation}
where $\varepsilon$ is the generalized minimal eigenvalue.

\subsection{\methodnameexc (\methodnameexcacronym)}
\label{sec:optimization}
In this section we introduce \methodnameexc (\methodnameexcacronym) to variationally optimize states consisting of sums of Slater determinants.
The method uses the fact that the subspace variational principle for excited states (see~\cref{sec-var-theorem-exc-states}) is equivalent to an effective ground-state optimization problem for a single state (see~\cref{sec-effective-hamiltonian-state}) that can be solved exactly for a linear ansatz (see~\cref{sec-linear-optimization}).

\subsubsection{The variational ansatz}

\label{subsubsec-var-ansatz-methods}
We consider here the case where each state $\ket{\ExcState^a}$, $a\in\{0,\dots,\nstates-1\}$, is a sum of generalized Hartree-Fock (GHF) Slater determinants
\begin{equation}\label{eq-ansatz-sum-det}
    \ket{\ExcState^{a}} = \sum_{I=1}^{\ndet} \ket{\SDet^{a I}},
\end{equation}
where each $\ket{\SDet^{aI}}$ is a Slater determinant and where, for simplicity, we assume a uniform number of determinants $\ndet$ for every state.
Each determinant is encoded as
\begin{equation}
    \label{eq:Slater_determinant_orb_creation}
    \ket*{\SDet^{a I}}= \left(\sum_{\mu_1=1}^{\norb}\SDeTen^{a I}_{1\mu_1} \hat c^\dagger_{\mu_1}\right)\dots \left(\sum_{\mu_\nele=1}^{\norb}\SDeTen^{a I}_{\nele\mu_{\nele}} \hat c^\dagger_{\mu_{\nele}}\right)\ket{\vacuum},
\end{equation}
where $\norb$ is the number of basis-set orbitals, $\nele$ is the number of electrons, and $\ket{\vacuum}$ denotes the vacuum.

We defer the discussion of a specialization to unrestricted Hartree–Fock (UHF) Slater-determinants, used for the simulations in this work, to \cref{sec:uhf_ansatz}.

\subsubsection{Overview of the main idea}
The crucial observation at the basis of the \methodnameexcacronym technique is the realization that we can optimize an orbital in each determinant for a single state exactly and efficiently. This is based on the following properties:
\begin{enumerate}[label=(\roman*)]\item The ansatz for the $a$-th state defined in~\cref{eq-ansatz-sum-det} and~\cref{eq:Slater_determinant_orb_creation} is linear in $\SDeTen^{aI}_{j \mu_j}$ at fixed $j$.
More precisely, we define our optimization parameters $\bm \theta$ to be  
    \begin{equation}
    \bm\theta \coloneqq \left(\SDeTen^{a I}_{j \mu}\right)^{I\in\{1,\dots\ndet\}}_{\mu\in\{1,\dots,\norb\}},
\end{equation}
so $\bm \theta\in \mathbb{C}^{\ndet\times \norb}$.
This fact is also used in the ground-state \methodnameacronym method of Ref.~\cite{paper1}.
\item \cref{sec-linear-optimization} shows that we can optimize exactly a linear ansatz for the $a$-th state by solving a generalized eigenvalue problem.
\item We can apply the efficient tensor-contractions techniques developed in the \methodnameacronym paper, see Ref.~\cite{paper1}, to compute the matrices defining the generalized eigenvalue problem of~\cref{eq:tr_quad_form}.
\end{enumerate}
 The strategy of \methodnameexcacronym is therefore the following: at each step of the loop, until convergence: 
\begin{enumerate*}[label=(\roman*)]\item (Optional) We rotate the orbitals of each determinant of each state. This does not change the state, but could help to improve convergence;\item  We choose a state $a$ and an orbital $j$ to optimize; \item We optimize exactly the subspace loss function as function of the state $a$ and the orbital $j$ by solving a generalized eigenvalue problem. 
\end{enumerate*}

The complete algorithm is described in the next section.

\subsubsection{The \methodnameexcacronym optimization loop}\label{subsub-exidosloop}

\begin{algorithm}[H]
\caption{Sketch of the \methodnameexcacronym algorithm}
\label{alg:algo}
\begin{algorithmic}[1]
\State \texttt{init}: Initialize Slater determinants $\ket{\Phi^{bI}}$ with $\SDeTen^{bI}_{i\mu}\in\mathbb{C}$
\While{not converged}
    \State \texttt{gauge\_transform}: (Optional) Mix all orbitals:
    $\SDeTen^{bI}_{i\mu} \rightarrow \sum_{k=1}^{\nele} V^{bI}_{ik}\, \SDeTen^{bI}_{k\mu}$, with $V^{bI}\in SL(\nele,\mathbb{C})$ 
    \State \texttt{choose\_state\_orb}: Choose a state $a$ and an orbital $j$ to optimize. Introduce  $    \bm\theta \coloneqq \left(\SDeTen^{a I}_{j \mu}\right)^{I\in\{1,\dots\ndet\}}_{\mu\in\{1,\dots,\norb\}}$
    \State \texttt{build\_gev\_matrices}: Calculate $\HQuad$ and $\SQuad$ (\cref{{eq:squad_def,eq:hquad_def}}) for the chosen state $a$ and the chosen orbital $j$
    \State \texttt{solve\_gev}: Solve the GEV $\HQuad \bm \theta = \epsilon \SQuad \bm \theta$
    \State \texttt{update\_state\_orb}: Use the solution of the GEV, $\bm \theta$, to update $\SDeTen^{aI}_{j\mu}\rightarrow\theta^I_\mu$ for $I\in\{1,\dots,\ndet\}$ and ${\mu\in\{1,\dots,\norb\}}$
\EndWhile
\State \texttt{post\_proc}: Call either \texttt{naive\_post} (see~\cref{alg:post1}) or \texttt{impr\_post} (see~\cref{alg:post2}) 
\State \texttt{symm\_diag}: (Optional) Call \texttt{symm\_diag} (\cref{alg:deg_diag_sub} in 
\cref{appendix:diagonalization})
\end{algorithmic}
\end{algorithm}

We sketch the \methodnameexcacronym method in~\cref{alg:algo}. In this section we give a high-level description of each step of this technique.

\paragraph*{\textup{\texttt{init}}:}
We initialize the coefficients $\SDeTen^{aI}_{j\mu}\in\mathbb{C}$ of the determinants, representing the set of states $\{\ket{\ExcState^a}\}_{a=1}^{\nstates}$. Generally, we draw all the coefficients independently from a Gaussian distribution. If an initial guess from a previous simulation exists, we only set stochastically the coefficients for which an initial guess is not available.

\paragraph*{\textup{\texttt{gauge\_transform}} (optional):}
Each state $b$ is invariant under a ``gauge transformation'' that mixes the orbitals of each determinant $I$: ${\tilde{\SDeTen}}^{bI}_{i\mu} = \sum_{k=1}^{\nele} V^{bI}_{ik}\, \SDeTen^{bI}_{k\mu}$, where $V^{bI}\in SL(\nele,\mathbb{C})$. This gauge transformation was also used in the \methodnameacronym algorithm of Ref.~\cite{paper1}. The independent mixing of the orbitals of each determinant can improve the convergence properties of the algorithm. Accordingly, we always use \texttt{gauge\_transform} in this work, but we note that it is not strictly needed for \methodnameexcacronym convergence. We choose to draw $V^{bI}$ from a uniform probability distribution, independently for each state $b$ and each determinant $I$.

\paragraph*{\textup{\texttt{choose\_state\_orb}}:}
We choose a state $a$ and an orbital $j$ to optimize. Regarding the choice of the state, we usually sample $a$ uniformly from the set $\{0, \dots, \nstates-1\}$. Other choices, such as sweeps, are possible. 
As we always use the optional  \texttt{gauge\_transform} subroutine in our numerical results, the choice of the orbital $j$ is superfluous as a permutation is a special case of the gauge transformation, and correspondingly we always choose to optimize the last orbital, $j=\nele$. The choice of the state $a$ and the orbital $j$ defines the optimization parameters $\bm\theta$ for the step of the optimization loop, $    \bm\theta \coloneqq \left(\SDeTen^{a I}_{j \mu}\right)_{I\in{1,\dots\ndet},\mu\in\{1,\dots,\norb\}}$.
\paragraph*{\textup{\texttt{build\_gev\_matrices}}:}
We construct the effective overlap and  Hamiltonian matrices $\SQuad$ and $\HQuad$ matrices for the chosen state $a$ and the chosen orbital $j$ according to \cref{eq:squad_def,eq:hquad_def}, using the matrix elements from \cref{eq:mel_grad,eq:mel_hess}.
In order to apply the equations from \cref{sec-linear-optimization}, we reorder the states in such a way that state $a$ becomes the first state $0$.
$\SQuad$ and $\HQuad$ are made up from the matrix elements in \cref{eq:squad_def,eq:hquad_def}.
They can be computed efficiently using the tensor-contraction strategy we introduced in the ``Efficient calculation of the effective matrices'' section of Ref.~\cite{paper1}.

\paragraph*{\textup{\texttt{solve\_gev}}:}
We solve the generalized eigenvalue problem $\HQuad \bm \theta = \varepsilon\, \SQuad \bm\theta$ (\cref{eq:gev}) to find the optimal value for the optimization parameters $\bm \theta$.
The matrices $\HQuad$ and $\SQuad$ are singular, with the nullspace being spanned by  orbitals different from the chosen $j$ of each determinant $I$ of the chosen state $a$~\cite{paper1}.
For solving the generalized eigenvalue problem \cref{eq:gev}, we first project out the nullspace, use a Cholesky decomposition to convert it to a normal eigenvalue problem and use Lanczos to find the eigenvector associated with the lowest eigenvalue. See \cref{appendix:numdetails} for details.

\paragraph*{\textup{\texttt{update\_state\_orb}}:}
With the solution of \cref{eq:gev}, $\bm \theta$, provided by the \texttt{solve\_gev} routine, we update the coefficients of orbital $j$ of all the determinants $I$ in state $a$ accordingly.

\paragraph*{\textup{\texttt{post\_proc}}:}
After a sufficient number of iterative steps, the algorithm will have optimized all the orbitals $j$ of all the states $a$, resulting in an optimized set of states $\{\ket{\ExcState^a}\}_{a=0}^{\nstates-1}$. In a loose sense, all orbitals are partially optimized at each step of the optimization loop thanks to the \texttt{gauge\_transform} subroutine that mixes all orbitals.
What is left is to find the actual expression for the variational eigenstates and their energies, which we refer to as postprocessing. We discuss in detail two alternative postprocessing techniques in the next section.

\subsubsection{Postprocessing}
\label{sec:post}
\begin{algorithm}[H]
\caption{\texttt{naive\_post}: Naive Postprocessing}
\label{alg:post1}
\begin{algorithmic}[1]
\State Build $H$ and $S$ (\cref{eq:mel_S_H})
\State Solve the GEV  $H v = \varepsilon S v$ computing all eigenpairs
\end{algorithmic}
\end{algorithm}

\begin{algorithm}[H]
\caption{\texttt{impr\_post}: Improved Postprocessing}
\label{alg:post2}
\begin{algorithmic}[1]
\State Build $H_{aI,bJ} \coloneq \mel{\SDet^{aI}}{\hat H}{\SDet^{bj}}$ and $S_{aI,bJ} \coloneq \braket{\SDet^{aI}}{\SDet^{bj}}$
\State Solve the GEV $H\, v = \varepsilon\, S v$ for the lowest $\nstates$ eigenpairs 
\end{algorithmic}
\end{algorithm}

\paragraph*{\textup{\texttt{naive\_post}}:}
We build the overlap and Hamiltonian matrix \cref{eq:mel_S_H} in the subspace spanned by the optimized states 
$\{\ket{\ExcState^a}\}_{a=1}^{\nstates}$
and solve the generalized eigenvalue problem \cref{eq:eigval_problem_var_principle} directly, obtaining the approximate eigenstates \cref{eq:gamma_states} as linear combinations of the $\ket{\ExcState^a}$, with energy \cref{eq:gamma_energy}.

\paragraph*{\textup{\texttt{impr\_post}}:}

In the \texttt{naive\_post} approach we keep the linear combinations of the determinants within each state $\ket{\ExcState^a}$ fixed. However, by exploiting the structure of the ansatz and relaxing them, we can get improved estimates of the eigenstates.
By directly diagonalizing the Hamiltonian in the subspace spanned by all the determinants $\ket{\SDet^{aI}}$ for $I\in\{1,\dots, \ndet\}$ and $a\in\{0,\dots,\nstates-1\}$, we obtain a set of $\ndet\cdot \nstates$ orthogonal states with $\ndet\cdot \nstates$ determinants each, approximating the $\ndet\cdot \nstates$ lowest eigenstates.
From this set, we select the first $\nstates$ states to obtain an improved approximation of the excited states.
The variational principle we reviewed in \cref{sec:var_principle} is preserved, since we can view this procedure as $\nstates\cdot\ndet$ states with $1$ determinant each, where we discard all but the lowest $\nstates$ ones.
This procedure typically comes at little extra cost, since, due to the prefactors, the dominant part are the matrix elements between all pairs of determinants which are the same, and the solution of the larger eigenvalue problem is negligible.
Unless otherwise noted, we accordingly always use the \texttt{impr\_post} procedure in this work.

\paragraph*{\textup{\texttt{symm\_diag}}:}
\methodnameexcacronym, approximately diagonalizes the Hamiltonian at low energy and therefore provides a set of approximate $\hat H$ eigenvectors.
A symmetry of the Hamiltonian is an operator $\hat \Omega$ that commutes with it, $[\hat H,\hat \Omega]=0$. Therefore, there exists a basis in which both $\hat H$ and $\hat \Omega$ are diagonal.
In some cases, the $\hat H$ eigenvectors are degenerate. Numerically this appears as two or more eigenvectors that are very close in energy. In such cases, we introduce a tolerance to identify approximate degeneracies, and then diagonalize $\hat \Omega$ within the subspace spanned by the corresponding (approximately degenerate) eigenvectors of $\hat H$. This procedure is presented in more detail in \cref{alg:deg_diag_sub} of \cref{appendix:diagonalization}.
We remark that the variational principle from \cref{sec:var_principle} remains conserved for the average energy of the almost-degenerate states after these additional diagonalizations.

\subsubsection{Additional remarks}
 For a very large number of determinants $\ndet$, it could be necessary to decrease the size of the GEV in the \texttt{solve\_gev} subroutine by optimizing only a subset of the determinants. As an identical strategy is used for the EIDOS method, we refer to to Ref.\cite{paper1}, Supplementary Note 3, for details.

We note that for the special case of a single state, $\nstates=1$, \methodnameexcacronym reduces to the \methodnameacronym algorithm for ground-state optimization introduced in Ref.~\cite{paper1}, as expected. We also want to point out here the proposal to optimize a bosonic ground-state wave function written as a sum of permanents~\cite{Zhang2022_boson_permanent_quadratic_opt_a,Que2025_boson_permanent_quadratic_opt_b}. Similarly to \methodnameacronym, the bosonic technique uses the fact that the ground-state energy is quadratic when viewed as a function of one ``orbital'' in each permanent; however, contrarily to what happens in the fermionic case, no gauge transformation can be applied to permanents, the efficient \methodnameacronym tensor-contraction formulas are not immediately translatable to the permanent case, and, most importantly, the numerical cost to solve the generalized eigenvalue problem for a sum-of-permanents ansatz increases exponentially with the number of bosons.

\subsection{Computational scaling}
\label{sec:scaling}
As discussed in~\cref{subsub-exidosloop}, in the \texttt{build\_gev\_matrices} subroutine we compute the matrix elements defined in \cref{eq:mel_S_H,eq:mel_grad,eq:mel_hess} for the sum of Slater determinant ansatz of~\cref{eq:Slater_determinant_orb_creation}.
This has an asymptotic cost of $O(\norb^4)$ for each pair of determinants at fixed number of electrons $\nele$ (see Ref.~\cite{paper1}).
We cache $\SES$ and $\HES$ between iterations, updating the row and column corresponding to the state whose parameters are updated from the solution of \cref{eq:gev}.
In this way the number of new matrix elements we need to compute for $\SRed$ and $\HRed$ at every iteration is linear in the number of target states $\nstates$.
Computing $\SES,\HES$ themselves scales cubically in $\nstates$, due to the $\SRed^{-1}$ term. 
However, since typically the number of states we study is much smaller than the basis set size $\norb$, $\nstates \ll {\norb}^4$, the linear term in $\nstates$ coming from the calculation of $\mathcal T$ and $\mathcal G$ dominates. Therefore, the overall asymptotic scaling of a single iteration of the \methodnameexcacronym algorithm is linear in the number of states $K$.
Additionally, we remark that it is possible to efficiently compute matrix elements of $\hat H^2$ for our ansatz \cref{eq:Slater_determinant_orb_creation}, and thus the variance of the Hamiltonian (see Supplementary Note 1 of Ref. \cite{paper1}).

\subsection{Simulation Setup}
\label{sec:simulation_setup}
\subsubsection{The model}
We work with the second-quantized formulation of the electronic molecular Hamiltonian in a finite basis set of size $\norb$, given by
\begin{equation}
\label{eq:hamiltonian}
    \hat H
    = \sum_{\substack{\mu\nu\\\sigma}} h_{\mu\nu}\, \hat c_{i\sigma}^\dagger \hat c_{j\sigma}
    + \frac{1}{2} \sum_{\substack{\mu\nu\xi\zeta\\\sigma\sigma^\prime}} h_{\mu\nu\xi\zeta}\, \hat c_{\mu\sigma}^\dagger \hat c_{\xi\sigma^\prime}^\dagger \hat c_{\zeta\sigma^\prime} \hat c_{\nu\sigma}
    + E_{\textrm{nuc}},
\end{equation}
where $\mu,\nu,\xi,\zeta \in \{1,\dots, \norb\}$ and $\sigma,\sigma^\prime \in \{\uparrow, \downarrow\}$, and $E_{\textrm{nuc}}\in\mathbb{R}$ is the nuclear-nuclear energy.

\subsubsection{The sum-of-UHF-determinants ansatz for the states}
\label{sec:uhf_ansatz}
In~\cref{subsubsec-var-ansatz-methods} we have considered the variational ansatz to be  a sum of GHF Slater determinants (see~\cref{eq:Slater_determinant_orb_creation}).
For the numerical implementation, in the context of this work, it is computationally convenient to use a specialized form of this ansatz based on unrestricted Hartree-Fock (UHF) Slater determinants $\ket{\SDet^{aI}}$, $I\in\{1,\dots,\ndet\}$, ${a\in\{0,\dots,\nstates-1\}}$, defined by
\begin{equation}
    \label{eq:Slater_determinant_orb_creation_uhf}
    \begin{split}
    \ket*{\SDet^{aI}}&=
    \left(\sum_{\mu_1=1}^{\norb}\SDeTen^{aI}_{1\mu_1{\uparrow}}\; \hat c^\dagger_{\mu_1 \uparrow}\right)\dots \left(\sum_{\mu_{\nele_\uparrow}=1}^{\norb}\SDeTen^{aI}_{\nele_\uparrow\mu_{\nele_\uparrow}{\uparrow}}\; \hat c^\dagger_{\mu_{\nele_\uparrow}\uparrow}\right)\times\\
&\times           \left(\sum_{\nu_1=1}^{\norb}\SDeTen^{aI}_{1\nu_1{\downarrow}}\; \hat c^\dagger_{\nu_1 \downarrow}\right)\dots \left(\sum_{\nu_{\nele_\downarrow}=1}^{\norb}\SDeTen^{aI}_{\nele_\downarrow\nu_{\nele_\downarrow}{\downarrow}}\; \hat c^\dagger_{\nu_{\nele_\downarrow}\downarrow}\right) \ket{\vacuum}
    \end{split}
\end{equation}
where $n_\uparrow$ ($n_\downarrow$) is the number of spin-up (spin-down) electrons, $\norb$ is the number of basis-set orbitals and the unconstrained variational parameters $\SDeTen_{j\mu\sigma}^{aI}\in\mathbb{C}$ represent a set of $\nstates$ states with $\ndet$ determinants each. 

Each UHF determinant $\ket*{\SDet^{aI}}$ is by construction an eigenstate of $\hat S_z$ with fixed eigenvalue ${\ev*{\hat S_z} = \frac{n_\uparrow-n_\downarrow}{2} \eqqcolon M}$ and an eigenstate of the total electron number with eigenvalue $n_\uparrow+n_\downarrow$. 

The use of UHF-type determinants allows to speed up the \methodnameexcacronym algorithm we introduce in this work by halving the effective number of orbitals  when compared to the use of GHF-type determinants. The only modification needed in the \texttt{choose\_state\_orb} subroutine of~\cref{subsub-exidosloop} is choosing whether to optimize the spin-up or spin-down part of each determinant at every step of the iterative algorithm (see also section C in the supplementary material of Ref.~\cite{paper1} for  details).

Unless otherwise noted, the energies provided are obtained by using \texttt{impr\_post} (\cref{alg:post2}) as outlined in \cref{sec:post}.

The \methodnameexcacronym method approximately diagonalizes $\hat H$ at low energy. When several of the approximate eigenstates are numerically close to being degenerate, we diagonalize the symmetry operators which commute with the Hamiltonian within each degenerate subspace to obtain approximate symmetry eigenstates using the procedure described in \cref{appendix:diagonalization}.

\FloatBarrier

\section*{Acknowledgements}
We thank E. Monino for discussions. 

\FloatBarrier

\bibliography{main.bib}

\clearpage
\appendix
\onecolumngrid

\begin{center}
\textbf{\large Appendix}\\
\end{center}
\crefalias{section}{appendix}

\section{Derivation of the effective operators}
\label{appendix:derivation_operators}
Let $\mathcal{U} = \{\ket{\ExcState^a}\}_{a=0}^{\nstates-1}$ be a set of variational states drawn from independent
variational manifolds $\ket{\ExcState^a} \in \mathcal{M}_a$. Let $\mathcal V = \mathrm{span}\{\ket{\ExcState^a}\} \subseteq \mathcal{H}$
be the space they span, $\hat P$ the corresponding projector ($\mathcal V = \hat P\mathcal{H}$) and $\mathcal{L}$ the cost function
\begin{equation}
      \mathcal{L} = \mathrm{Tr}(\hat P \hat H)
\end{equation}
where $\hat H$ is the system Hamiltonian. Both the projector and the cost function
can be explicitly written in terms of the states $\mathcal{U}$. The projector is
\begin{equation}
  \hat P = \sum_{a,b=0}^{\nstates-1}  \left(S^{-1}\right)_{ab} \dyad{\ExcState^{a}}{\ExcState^{b}},
\end{equation}
where $S_{ab} = \braket{\ExcState^a}{\ExcState^b}$ for $a,b = 0,\ldots,{\nstates-1}$, while the cost function can be
rearranged as
\begin{equation}
\begin{aligned}
\label{eq:tr_op_mat_equiv}
  \mathcal{L}
  &= \mathrm{Tr}(\hat P \hat H)\\
  &= \sum_{n \in N} \bra{n} \hat H \hat P \ket{n}\\
  &= \sum_{n \in N} \sum_{a,b=0}^{\nstates-1} \left(S^{-1}\right)_{ab} \mel{n}{\hat H}{\ExcState^a} \braket{\ExcState^b}{n} \\
  &= \sum_{a,b=0}^{\nstates-1} \left(S^{-1}\right)_{ab} \mel{\ExcState^b}{\hat H}{\ExcState^a}\\
  &\equiv \sum_{a,b=0}^{\nstates-1} \left(S^{-1}\right)_{ab} H_{ba}\\
  &= \mathrm{tr}\!\left(S^{-1} H\right)
\end{aligned}
\end{equation}
upon introducing an arbitrary, auxiliary complete orthonormal set of states
$\{\ket{n}\}_{n \in N}$.
Here $H_{ab} = \mel{\ExcState^a}{\hat H}{\ExcState^b}$ and $S_{ab} = \braket{\ExcState^a}{\ExcState^b}$ are the
$\mathcal{U}$-representations of $\hat H$ and $\hat{\mathbf{1}}$. Note that we have used the capital symbol
``$\mathrm{Tr}$'' for the operator trace, and the normal-case symbol ``$\mathrm{tr}$'' for the matrix trace.

Let us now consider the partial optimization problem involving a single state,
say $a=0$.
To this end, let $\tilde{\mathcal{U}} = \{\ket{\ExcState^a}\}_{a=1}^{\nstates-1}$ be the fixed states,
$\tilde{\mathcal V} = \mathrm{span}(\tilde{\mathcal{U}})$ the space they span and ${\hat P}_{\tilde {\mathcal{V}}}$ the corresponding projector, whose explicit form is given in \cref{eq:def_pitilde} below.
Then, we can decompose $\hat P$ as follows
\begin{equation}
  \hat P = {\hat P}_{\tilde {\mathcal{V}}} + \hat P_0
\end{equation}

where the rank-1 projector $\hat P_0$ reads as
\begin{equation}
  \hat P_0 = \frac{\hat{\mathcal{S}}\dyad{\ExcState^{0}}{\ExcState^{0}}\hat{\mathcal{S}}}{\mel{\ExcState^{0}}{\hat{\mathcal{S}}}{\ExcState^{0}}}
\end{equation}
with
\begin{equation}
\label{eq:def_s_op_full}
  \hat{\mathcal{S}} = \hat{\mathbf{1}} - {\hat P}_{\tilde {\mathcal{V}}}
\end{equation}
This follows from the fact that $\hat{\mathcal{S}} \ket{\ExcState^{0}}$ lies in $\mathcal V$ and it is orthogonal to $\tilde{\mathcal V}$, by
construction. 
Using the above decomposition we find
\begin{equation}
  \mathcal{L} = \mathrm{Tr}({\hat P}_{\tilde {\mathcal{V}}} \hat H) + \mathrm{Tr}(\hat P_0 \hat H)
\end{equation}

where the second term takes the form of a Rayleigh quotient $
  \mathrm{Tr}(\hat P_0 \hat H) = \frac{\mel{\ExcState^{0}}{\hat{\mathcal{S}} \hat H \hat{\mathcal{S}}}{\ExcState^{0}}}{\mel{\ExcState^{0}}{\hat{\mathcal{S}} }{\ExcState^{0}}}
$.
Equivalently,
by multiplying and dividing the first term with $\mel*{\ExcState^{0}}{\hat{\mathcal{S}} }{\ExcState^{0}}$, we have that
\begin{equation}
  \mathcal{L} = \frac{\mel{\ExcState^{0}}{\hat{\mathcal H}}{\ExcState^{0}}}{\mel{\ExcState^{0}}{ \hat{\mathcal{S}}}{\ExcState^{0}}}
\end{equation}

where $\hat{\mathcal{H}}$ is
\begin{equation}
\label{eq:h_op_def1}
  \hat{\mathcal{H}} = \hat{\mathcal{S}}\,\mathrm{Tr}({\hat P}_{\tilde {\mathcal{V}}} \hat H) + \hat{\mathcal{S}} \hat H \hat{\mathcal{S}}
\end{equation}

and does not depend on $\ket{\ExcState^{0}}$, since both $\hat{\mathcal{S}}$ and ${\hat P}_{\tilde {\mathcal{V}}}$ depend only on the fixed set
$\tilde{\mathcal{U}}$.

\subsubsection*{Explicit expression of the $\hat{\mathcal{S}}$ and $\hat{\mathcal{H}}$ operators}

The $\hat{\mathcal{S}}$ and  $\hat{\mathcal{H}}$ operators can be given in more explicit form in terms of the fixed elements
$\tilde{\mathcal{U}} = \{\ket{\ExcState^a}\}_{a=1}^{\nstates-1}$.
For the $\hat{\mathcal{S}}= \hat {\mathbf{1}} - {\hat P}_{\tilde {\mathcal{V}}}$ projector we need the explicit expresion for the projector onto $\tilde{V}$, given by
\begin{equation}
\label{eq:def_pitilde}
{\hat P}_{\tilde {\mathcal{V}}} = \sum_{a,b=1}^{\nstates-1}  \left(\tilde{S}^{-1}\right)_{ab} \dyad{\ExcState^a}{\ExcState^b}
\end{equation}

Here, $\tilde{S}$ is the $(\nstates{-}1)\times(\nstates{-}1)$ block of the overall overlap matrix that only
involves the states $\tilde{\mathcal{U}}$, i.e. the lower right block of
$
S \eqqcolon \pma{
s & \omega^\dagger \\
\omega & \tilde{S}}
$ where
$s = \braket{\ExcState^{0}}{\ExcState^{0}}$ and $\omega_a = \braket{\ExcState^{a}}{\ExcState^{0}}$ for $a \in \{1,\dots,{\nstates-1}\}$.

Consequently, for $\hat{\mathcal{H}}$,  following from \cref{eq:h_op_def1,eq:def_pitilde},
\begin{equation}
\label{eq:h_op_def_full}
\hat{\mathcal{H}} =
\hat H - {\hat P}_{\tilde {\mathcal{V}}} \hat H
-  \hat H  {\hat P}_{\tilde {\mathcal{V}}}
+ {\hat P}_{\tilde {\mathcal{V}}} \hat H {\hat P}_{\tilde {\mathcal{V}}}
+ f \hat{\mathcal{S}}
\end{equation}
where, by \cref{eq:tr_op_mat_equiv},
\begin{equation}
\label{eq:def_f}
  f = \mathrm{Tr}({\hat P}_{\tilde {\mathcal{V}}} \hat H) = \mathrm{tr}\!\left(\tilde{S}^{-1}\tilde{H}\right),
\end{equation}
and $\tilde{H}$ is the lower-right $(\nstates{-}1)\times(\nstates{-}1)$ block of the matrix $
  H \eqqcolon \pma{
  h & \tau^\dagger \\ 
  \tau & \tilde{H} }$
where
$h = \mel{\ExcState^{0}}{\hat H}{\ExcState^{0}}$ and $\tau_a = \mel{\ExcState^a}{\hat H}{\ExcState^{0}}$ for $a \in \{1,\dots,{\nstates-1}\}$.

\section{Derivation of the effective matrices}
\label{appendix:derivation_tensors}

We derive the explicit expressions for the matrix elements
${\SQuad_{pq} =  \mel{\ExcState^{0}_p}{\hat S^{(0)}}{\ExcState^{0}_q}}$ and
${\HQuad_{pq} =  \mel{\ExcState^{0}_p}{\hat H^{(0)}}{\ExcState^{0}_q}}$.

Using \cref{eq:def_pitilde} we have that
\begin{equation}
\mel{\ExcState^{0}_p}{{\hat P}_{\tilde {\mathcal{V}}}}{\ExcState^{0}_q} =
\sum_{a,b=1}^{\nstates-1}  \left(\tilde{S}^{-1}\right)_{ab}
\underbrace{\braket{\ExcState^{0}_p}{\ExcState^{a}}}_{=\mathcal T_{ap}^\dagger}
\underbrace{\braket{\ExcState^{b}}{\ExcState^{0}_q}}_{=\mathcal T_{bq}}
\end{equation}
as well as
\begin{equation}
\mel{\ExcState^{0}_p}{{\hat P}_{\tilde {\mathcal{V}}} \hat H }{\ExcState^{0}_q} =
\sum_{a,b=1}^{\nstates-1}  \left(\tilde{S}^{-1}\right)_{ab}
\underbrace{\braket{\ExcState^{0}_p}{\ExcState^{a}}}_{=\mathcal T_{ap}^\dagger}
\underbrace{\mel{\ExcState^{b}}{\hat H}{\ExcState^{0}_q}}_{=\mathcal G_{bq}}.
\end{equation}
We further find
\begin{equation}
\label{eq:php}
{\hat P}_{\tilde {\mathcal{V}}}\hat H  {\hat P}_{\tilde {\mathcal{V}}} = \sum_{a,b,c,d=1}^{\nstates-1}  \left(\tilde{S}^{-1}\right)_{ab} \left(\tilde H\right)_{bc}
\left(\tilde{S}^{-1}\right)_{cd}
\dyad{\ExcState^{a}}{\ExcState^{d}}
= \sum_{a,b=1}^{\nstates-1}  \left(\tilde{S}^{-1} \tilde H\tilde{S}^{-1}\right)_{ab}
\dyad{\ExcState^b}{\ExcState^a}
\end{equation}
and thus
\begin{equation}
\mel{\ExcState^{0}_p}{{\hat P}_{\tilde {\mathcal{V}}} \hat H  {\hat P}_{\tilde {\mathcal{V}}} }{\ExcState^{0}_q} =
\sum_{a,b=1}^{\nstates-1}  \left(\tilde{S}^{-1} \tilde H\tilde{S}^{-1}\right)_{ab}
\underbrace{\braket{\ExcState^{0}_p}{\ExcState^a}}_{=\mathcal T_{ap}^\dagger}
\underbrace{\braket{\ExcState^b}{\ExcState^{0}_q}}_{=\mathcal T_{bq}}.
\end{equation}

Then, using \cref{eq:def_s_op_full} we have that
\begin{equation}
{\SQuad_{pq} =  \mel{\ExcState^{0}_p}{\hat S^{(0)}}{\ExcState^{0}_q}} =
\underbrace{\braket{\ExcState^{0}_p}{\ExcState^{0}_q}}_{=\tilde{\mathcal S}_{pq}}
- \sum_{a,b=1}^{\nstates-1}  \mathcal T_{ap}^\dagger \left(\tilde{S}^{-1}\right)_{ab}
\mathcal T_{bq}
\end{equation}
and, by \cref{eq:h_op_def_full},
\begin{equation}
{\HQuad_{pq} =  \mel{\ExcState^{0}_p}{\hat H^{(0)}}{\ExcState^{0}_q}} =
\underbrace{\mel{\ExcState^{0}_p}{\hat H}{\ExcState^{0}_q}}_{=\tilde{\mathcal H}_{pq}}
-\sum_{a,b=1}^{\nstates-1} \mathcal T_{ap}^\dagger \left(\tilde{S}^{-1}\right)_{ab}\mathcal G_{bq}
-\sum_{a,b=1}^{\nstates-1} \mathcal G_{ap}^\dagger \left(\tilde{S}^{-1}\right)_{ab}\mathcal T_{bq}
+ \sum_{a,b=1}^{\nstates-1}  \mathcal T_{ap}^\dagger \left(\tilde{S}^{-1} \tilde H\tilde{S}^{-1}\right)_{ab} \mathcal T_{bq}
+ f \SQuad_{pq}
\end{equation}
where $f$ is defined in \cref{eq:def_f}.

\section{Numerical Details}
\label{appendix:numdetails}
\subsection{Projecting out the nullspace and solving the generalized eigenvalue problem}
\label{appendix:nullspaceproj}

We numerically construct the projection matrix by diagonalizing each $\norb\times \norb$ block on the diagonal of $\tilde{\mathcal{S}}$.
Then we Cholesky-
decompose $\tilde{\mathcal{S}} = L L^\dagger$ and use Lanczos to find the lowest eigenvector of $C = L^{-1}  \tilde{\mathcal{H}} (L^\dagger)^{-1}$, transforming back at the end with $L$.
For Lanczos solver we can use the previous coefficients as initial guess to reduce the number of iterations.
\subsection{Numerical stability}
\label{appendix:numstability}
To ensure numerical stability we normalize all states $\ket{\ExcState^{a}}$, and for the calculation of the matrix elements we also normalize the orbitals of the individual determinants $\ket{\Phi_{ik}}$ and use a sufficiently large Krylov space to render the Lanczos solver numerically stable.

\section{Dipole moment and oscillator strentgh}
\label{appendix:dip_f}

We compute the dipole moment between states $k$ and $l$, given by
\begin{equation}
    \boldsymbol{\mu_{ab}} = -\frac{\mel{\ExcState^a}{\boldsymbol{\hat r}}{\ExcState^b}}{\norm{\Psi^b} \norm{\ExcState^a} }
\end{equation}
where $\boldsymbol{\hat r}$ is the second quantized representation of the position operator, which is a one-body operator .

From its components we can then calculate the dimensionless oscillator strength, given by
\begin{equation}
   f_{ab} = \frac{2}{3} \frac{m_e}{\hbar^2} (E_a - E_b)\,\norm{\boldsymbol{\mu_{ab}}}^2
\end{equation}
where $E_a,E_b$ are the energies of the two states and ${\hbar = m_e = 1}$ in atomic units.

\section{Determining the symmetry of a nearly-degenerate state }
\label{appendix:diagonalization}
We illustrate in this Section the algorithm we use to determine the symmetry of nearly-degenerate states.

\begin{algorithm}[H]
\caption{\texttt{symm\_diag}: Degenerate subspace diagonalization}
\label{alg:deg_diag_sub}
\begin{algorithmic}[1]
\Function{diagonalize\_subspace}{$\hat \Omega$, Q}
    \State Compute ${\omega_{ab} = \mel*{\ExcState^a}{\hat \Omega}{\ExcState^b}}$ for ${\ket*{\ExcState^a},\ket*{\ExcState^b}\in Q}$
    \State Compute ${\varsigma_{ab} = \braket*{\ExcState^a}{\ExcState^b}}$ for ${\ket*{\ExcState^a},\ket*{\ExcState^b}\in Q}$
    \State Fully solve $\omega c = \lambda \varsigma c$
    \State $\tilde{Q} = \{\sum_{b=1}^{|Q|} c_{ab} \ket{\ExcState^b}\}_{a=1}^{|Q|}$
    \State \Return $\tilde{Q}$
\EndFunction
\State
\Function{expect}{$\hat \Omega$, Q}
    \State \Return $\Big[\frac{\mel*{\ExcState^a}{\hat \Omega}{\ExcState^a}}{\braket*{\ExcState^a}{\ExcState^a}} \mathrm{for} \ket{\ExcState^a} \in  Q\Big]$
\EndFunction
\State 
\Function{groupby}{$\hat \Omega$,Q, tolerance}
    \State compute $\lambda$ = \Call{expect}{$\hat \Omega$, Q}
    \State \Return {a list of distinct subsets of states in $Q$, grouped such that the expectation values $\lambda$ of the states in each subset differ by at most tolerance}
\EndFunction

\State
\Function{diagonalize}{$\hat \Omega$, $\Theta$, Q}
    \If{$|\Theta|$ = 0}
        \State \Return \Call{diagonalize\_subspace}{$\hat \Omega$, $Q$}
    \EndIf
    \State $R = \{\}$

    \State $\hat \Omega^\prime$, $\Theta^\prime$ = $\Theta[0], \Theta[1:]$
    \For{$Q^\prime$ in \Call{groupby}{$\hat \Omega$, Q, tolerance} }
        \If{$|Q^\prime| > 1$}
           \State $Q^\prime$ = \Call{diagonalize\_subspace}{$\hat \Omega^\prime$, $Q^\prime$}
           \State $R = R\, \cup $  \Call{diagonalize}{$\hat \Omega^\prime$, $\Theta^\prime$, $Q^\prime$}
        \Else
            \State $R = R \cup Q^\prime$
        \EndIf
    \EndFor
    \State \Return $R$
\EndFunction
\State
\State \Call{diagonalize}{$\hat H$, [$\hat S^2$, \dots ], $\{\ket{\ExcState^a}\}_{a=1}^{\nstates}$, tolerance}
\end{algorithmic}
\end{algorithm}

\end{document}